\documentclass[doublecol,figures]{epl2}

\usepackage{amsmath}
\usepackage{amsfonts}
\usepackage{amssymb}
\usepackage{braket}
\usepackage{esint}

\DeclareMathOperator{\tr}{Tr}
\DeclareMathOperator{\im}{Im}
\DeclareMathOperator{\re}{Re}
\newcommand{\di}{\mathrm d}
\newcommand{\imi}{\mathrm{i}}

\newcommand{\resMat}{\mathbf{G}}

\title{From non-ergodic eigenvectors to local resolvent statistics and back: a random matrix perspective}
\shorttitle{Non-ergodic eigenvectors and local resolvent statistics} 

\author{Davide Facoetti\inst{1}\thanks{\email{davide.facoetti@kcl.ac.uk}} \and Pierpaolo Vivo\inst{1} \and Giulio Biroli\inst{2,3}}
\shortauthor{D. Facoetti \etal}

\institute{                    
  \inst{1} Department of Mathematics, 
  King's College London,
  Strand,
  London
  WC2R 2LS, UK\\
  \inst{2} IPhT, CEA/DSM-CNRS/URA 2306, CEA Saclay, F-91191 Gif-sur-Yvette Cedex, France\\
  \inst{3} Laboratoire de Physique Statistique, Ecole Normale Sup\'erieure, PSL Research University, Universit\'e Paris Diderot Sorbonne Paris-Cit\'e, Sorbonne Universit\'es UPMC Universit\'e Paris 06, CNRS, 24 rue Lhomond, 75005 Paris, France
}
\pacs{72.15.Rn}{Localization effects (Anderson or weak localization)}
\pacs{71.30.+h}{Metal-insulator transitions and other electronic transitions}
\pacs{05.30.-d}{Quantum statistical mechanics}

\abstract{We study the statistics of the local resolvent and non-ergodic properties of eigenvectors for a generalised Rosenzweig-Porter $N\times N$ random matrix model, undergoing two transitions separated by a delocalised non-ergodic phase.
Interpreting the model as the combination of on-site random energies $\{a_i\}$ and a structurally disordered
hopping, we found that each eigenstate is delocalised over $N^{2-\gamma}$ sites close in energy
$|a_j-a_i|\leq N^{1-\gamma}$ in agreement with Kravtsov \emph{et al.} 
(\textit{New. J. Phys.,} \textbf{17} (2015) 122002)
. 
Our other main result, obtained combining a recurrence relation for the resolvent matrix with insights from Dyson's Brownian motion, is to show that the properties of the non-ergodic delocalised phase can be probed 
studying the statistics of the local resolvent in a non-standard scaling limit.}

\begin{document}

\maketitle

The theoretical study of the non-equilibrium dynamics of isolated quantum systems has attracted
considerable interest in recent years, partly due to advances in experiments on trapped
ultra-cold atomic gases~\cite{Eisert2015}.
One of the most fundamental questions that arose is about the applicability of statistical mechanics
to quantum systems in presence of interactions and disorder, and the related Many-body localisation (MBL) transition~\cite{BaskoAleinerAltshuler}.
A system is in a MBL phase if taking interactions into account the many-body eigenstates are localised in Fock space.
The Fock space can be seen as a lattice with connectivity determined by two-body interactions.
Its structure is that of a very high dimensional lattice where loops are scarce, therefore reminiscent of the
Bethe lattice and random regular graphs (RRG).  Starting from the pioneering work~\cite{Altshuler1997}, Anderson localization on such lattices has been considered by many as a simplified case to study questions 
related to the MBL transition. It attracted a lot of attention recently~\cite{Biroli2012,DeLuca2014a,DeLuca2014b} because it could provide a test ground to analyse 
the delocalised non-ergodic or ``bad metal'' regime, which was predicted as an intermediate phase separating the fully delocalised and the MBL phases \cite{Altshuler1997,BaskoAleinerAltshuler}.
In this regime, eigenstates would be delocalised over a large number of configurations, but which only cover a very tiny fraction, vanishing for large system size, of the entire Fock space.
Although the existence of the MBL transition is now well established (at least for one dimensional systems) \cite{Nandkishore15}, the understanding of the delocalised non-ergodic phase is far from being completed. 
Some numerical results seem to indicate its presence in many-body systems~\cite{Pino2015,Goold2015} whereas its existence on Bethe lattices is under intense scrutiny and debated~\cite{Biroli2012,DeLuca2014b,Altshuler2016,Tikhonov2016a,Tikhonov2016b}. It is not clear at this stage whether the sub-diffusive behaviour 
found before the MBL transition \cite{BarLev2015,BarLev2016,Vosk2015,Potter2015,Bloch} is somehow related to it.

Given this state of the art, it is therefore useful to study simpler models  that could provide a playground to explore its nature and sharpen the questions about it. With this aim, the authors of Ref.~\cite{Kravtsov2015} proposed a random matrix model, the generalised Rosenzweig-Porter (GRP) model, as a relative of the RRG with random on-site energy. 
This was motivated by the known relation between the RRG and Gaussian invariant ensembles of random matrix theory~\cite{Oren2010}. They showed that the GRP model indeed undergoes two transitions:
a localisation transition and a separate ergodic transition, with an intermediate delocalised non-ergodic phase separating the two.

In this work we also focus on this model. Our aim is to further characterise the intermediate phase of the GRP model. 
We do so by applying a technique based on a recurrence relation for the resolvent matrix, and the Dyson Brownian motion.
Our main results consist in linking the statistics of the local resolvent to the properties of the mixed phase, and 
in combining these insights with the Brownian motion analysis to derive the scaling of the eigenstates. Besides the interest in the MBL context, our results are also relevant in other physical situations where 
quasi-delocalised states emerge, such as jamming \cite{Manning2015} and random matrix theory~\cite{Cao2016,bourgadeprivatecomm}. 

\section{The GRP random matrix model}
Following~\cite{Kravtsov2015}, we consider a generalisation of the Rosenzweig-Porter model~\cite{RosenzweigPorter}, with the Hamiltonian given by a $N\times N$ Hermitian matrix 
\begin{equation}\label{eq:GRP}
H = A + \frac{\mu}{N\strut^{\gamma/2}} V\ ,
\end{equation}
where $A$ is diagonal with real entries $a_i$, independently drawn from a probability density $p_A(a_i)$,
while $V$ belongs to the Gaussian unitary ensemble (GUE) with variance 1.
Drawing analogies with the RRG, the GUE matrix $V$ corresponds to the structural geometrical disorder, while A to the
on-site disorder.\footnote{The analogy with the RRG would suggest to choose $V$ real symmetric (GOE).
    We consider the unitary model because most of the literature focuses on it.
    Our conclusions apply to both versions of the model.}
The parameter $\gamma$ controls the relative magnitude of the two terms: it is a proxy for the strength of
the on-site disorder. \\
For $\gamma>2$, standard second order perturbation theory shows that the GUE term is a small regular perturbation 
(the perturbation of the eigenvalues is much smaller than their typical level spacing). As a consequence,  
the Hamiltonian is close to $A$ and hence eigenstates are completely localised.
Similarly, for $\gamma<1$ the first term is a small regular perturbation, hence 
the rotationally invariant $V$ term dominates, and 
the eigenstates are uniformly distributed on the unitary sphere, as for the GUE.
The value $\gamma=1$ was indeed shown to play a special role for the density of states~\cite{Brezin1996, Kunz1998,Pandey1995,Altland1997,Guhr1996}, which is given for $\gamma \leq 1$
by the Wigner semicircle distribution, and for $\gamma>1$ by $p_A$.
The value $\gamma=2$ instead governs the level statistics on the scale of the typical level spacing. 
Computing the spectral form factor, the unfolded two-point correlation function
was shown to be universal, \emph{i.e.} it does not depend on the specific form of $p_A$~\cite{Kunz1998,Kravtsov2015}.
It has the Wigner-Dyson form for $\gamma<2$, and Poisson for $\gamma>2$.
These results confirm that for $\gamma<1$ and $\gamma>2 $ the system is respectively fully delocalised and fully localised.
The regime $\gamma\in (1,2)$ instead is special: the density of states is given by $p_A$ and not by Wigner semi-circle but nevertheless the nearest neighbours level statistics has the Wigner-Dyson form.
As shown in~\cite{Kravtsov2015,Shukla2016} and discussed later on, this regime provides a simple example of a delocalised non-ergodic phase.

\section{The delocalised non-ergodic phase}
The authors of Ref.~\cite{Kravtsov2015} characterised the eigenstates for $\gamma\in (1,2)$,
finding the support set~\cite{DeLuca2014a} to be a fractal over $N^{D_1}=N^{2-\gamma}$ sites.
For large $N$, the eigenstates are supported over a large number of sites, so they are delocalised - but only over a fraction $\propto N^{1-\gamma}$ of all sites,
which tends to zero in the thermodynamic limit.

To study the spectral statistics we focus on the resolvent matrix
\begin{equation}
\resMat(z)=(z-H)^{-1}\ ,
\end{equation}
a standard tool of random matrix theory. It is a random complex function, which evaluated at
$z=\lambda-\imi\eta$ carries information about spectral quantities at energy $\lambda$,
on a scale $\eta$.
The (global) resolvent is $G(z)=\tr\resMat(z)/N$, while
the diagonal elements of $\resMat$ are known as the local resolvent.
The behaviour of $G(z)$ is completely featureless: in the large $N$ limit it converges to a non-fluctuating value of order one as long as $\eta>1/N$, as can be checked by using the spectral representation of $G(z)$.
The statistics of the local resolvent can be instead used as a tool to probe localisation
transitions. In general one focuses on its imaginary part  for $\eta \rightarrow 0$ \emph{after} the $N\rightarrow \infty$ limit is taken: in the localised phase the imaginary part vanishes whereas it remains finite in the delocalised phase, see, 
\emph{e.g.},~\cite{Tarquini2016}. As we shall show below, in order to probe the non-ergodic delocalised phase one instead needs to consider a different scaling limit and study how the statistics of the local resolvent evolves when $\eta$ goes to zero as $1/N^\delta$ 
for $\delta<1$.  

In the next section we derive the probability distribution for the local resolvent in the delocalised non-ergodic phase.
We then combine this with results from the Dyson Brownian motion analysis to get a 
complete picture of the non-ergodic delocalised phase.

\section{Local resolvent statistics and non-ergodic delocalised phase}
Using the block matrix inversion formula, it is possible to derive the equality in distribution
between random variables
\begin{equation}\label{eq:Grecurrence}
G_{00}^{(N+1)}(z)^{-1} \stackrel{d}{=} z -H_{00}
-\frac{\mu^2}{N^{\gamma}}\sum_{ij}G_{ij}^{(N)}(z)V_{0i}V_{j0}\ ,
\end{equation}
relating the probability distribution of the $N$+1-dimensional local resolvent to those of the
$N$-dimensional local resolvent and uncorrelated matrix elements.

We now argue that the sum in the RHS of~\eqref{eq:Grecurrence} is self-averaging with respect to the $V_{0i}$s and the $V_{ij}$s 
in the large $N$ limit. 
Defining the random variables $X=1/N\sum_i G_{ii}(z)|V_{0i}|^2$ and $Y=2/N \sum_{i<j}G_{ij}(z)V_{0i}V_{j0}$, eq.~\eqref{eq:Grecurrence}
can be written as 
\begin{equation}
 G_{00}^{(N+1)}(z)^{-1} \stackrel{d}{=} z
-H_{00}-\frac{\mu^2}{N^{\gamma-1}}(X+Y) \ .
\end{equation}
The expected value $\braket{G_{ii}(z)}$ is of order one and does not depend on $i$, hence $\braket{X}=\braket{G_{ii}(z)}=G(z)$. 
The variance of $X$ reads  $\frac{1}{N^2}\sum_{ij} \braket{G_{ii}(z)G_{jj}(z)  |V_{0i}|^2|V_{0j}|^2}_c$ and can be broken 
up into two terms, corresponding to $i\ne j$ and $i= j$ respectively. By perturbative arguments~\cite{Altland1997}, it can be shown that correlations between different components of the
local resolvent matrix go to zero for $N\rightarrow \infty$ thus 
implying that the former contribution to the sum vanishes. The latter can be similarly shown to be negligible 
using perturbation theory in $V$ for $\eta=1/N^\delta$ with $\delta<1$. As a consequence $X$ becomes a non-fluctuating quantity, equal to $G(z)$, in the large $N$ limit. Instead $Y$ can be neglected since its average is zero and arguments analogous to the ones above imply that its variance vanishes.  
\begin{figure*}
    \begin{center}
    \includegraphics[width=.95\textwidth]{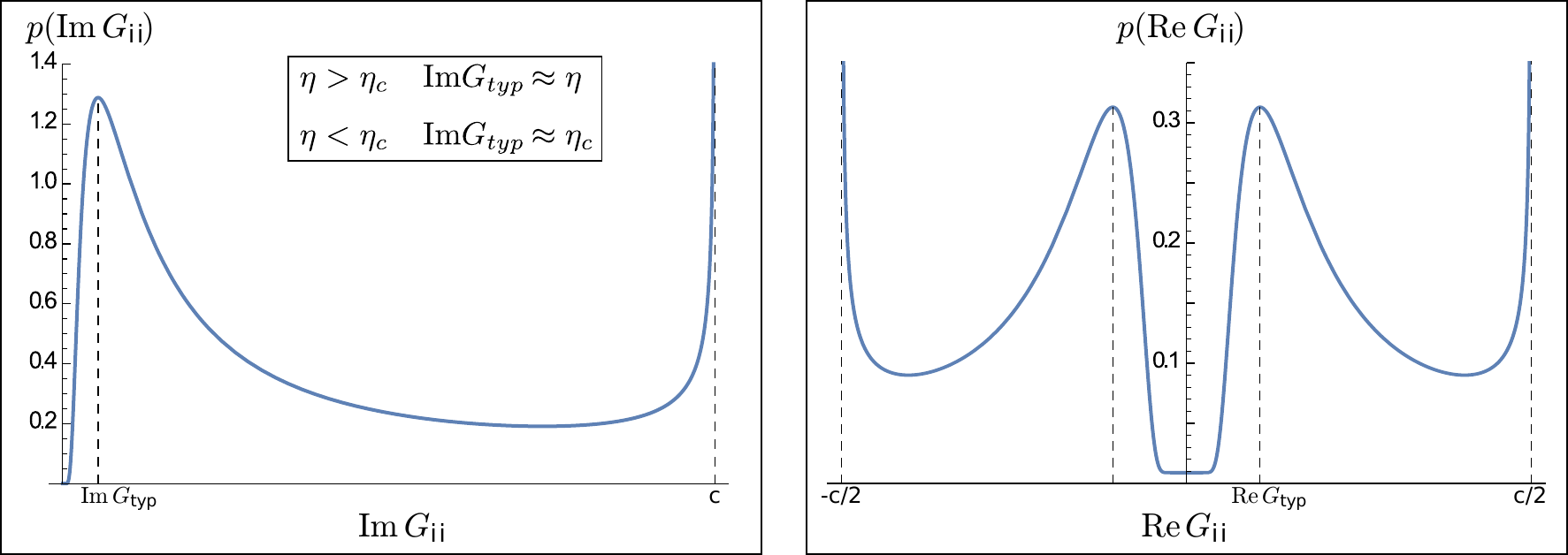}
    \caption{Probability distributions for the imaginary (left) and real (right) parts of the local
        resolvent $G_{ii}(z)$.
        The values $\lambda=0$, $\eta=0.4$ (left), $\eta=0.15$ (right)
        and a standard normal distribution for $p_A$ are used for the plots.
        For small $\eta>\eta_c$, $\im G_{typ}\propto \eta$, for $\eta<\eta_c$, $\im G_{typ}\propto \eta_c$ while $\re G_{typ}$ is of order one in both cases. The cutoffs are controlled by $c=1/\eta$ or
        $c=1/\eta_c$ respectively.}
    \label{fig.1}
    \end{center}
\end{figure*}
Therefore we obtain the relation
\begin{equation}\label{eq:Gtransitions}
  G_{00}(z)^{-1} \stackrel{d}{=} z - a - \frac{\mu^2}{N^{\gamma-1}} {G}(z) - \frac{\mu}{N^{\gamma/2}} V_{00}\ ,
\end{equation}
where all the randomness in the RHS comes from $a~\sim~ p_A(a)$ and $V_{00}\sim \mathcal{N}(0,1)$.
Since all diagonal elements of $\resMat$ are statistically equivalent, eq.~\eqref{eq:Grecurrence} establishes the distribution of $G_{ii}^{-1}$ for every $i$.
Looking at eq. (\ref{eq:Gtransitions}) one immediately realises that the values $\gamma=1$ and $\gamma=2$ 
play a special role. In our region of interest $\gamma\in (1,2)$,
the last term can be neglected. \\
By taking the average of the local resolvent and using eq. (\ref{eq:Gtransitions}), one finds that up to corrections small in $N$, the global resolvent is that of $A$,
\begin{equation}\label{eq:G3}
G(z)= \overline{G}(z)=\overline{G}_A(z)=
 \fint\frac{p_A(a)}{z-a}\di a\ ,
\end{equation}
where $\fint$ indicates the Cauchy principal value of the integral. Now that $G(z)$ is known, and is determined from 
$p_A$, the distribution of $G_{ii}(z)$ can be obtained. \\
We will study separately the real and imaginary parts of $G_{ii}(z)$ at $z=\lambda-\imi\eta$,
taking the large $N$ limit with $\eta$ either fixed, or scaling as $\eta\approx N^{-\delta}$ ($\delta<1$).
Up to terms that can be neglected in both cases, we derive from~\eqref{eq:G3} the expressions
\begin{align}\label{eq:reg1}
\re G_{ii}(z) &=\frac{\lambda-a_i}
{(\lambda-a_i)^2+\left(\eta+\frac{\mu^2}{N^{\gamma-1}}\im\overline{G}\right)^2}\ ,\\
 \im G_{ii}(z) &=\frac{\eta+\frac{\mu^2}{N^{\gamma-1}}\im\overline{G}(z)}
 {(\lambda-a_i)^2+\left(\eta+\frac{\mu^2}{N^{\gamma-1}}\im\overline{G}\right)^2}\ .\label{eq:img1}
\end{align}
Let us first focus on the usual scaling limit that corresponds to the large $N$ limit with $\eta$ small but fixed. The second term in the numerator of \eqref{eq:img1} then is subleading. Neglecting it, we obtain the distribution

\begin{equation}\label{eq:img-pdf}
p_{\im G}(x) =\left[\varphi_+(x)+\varphi_-(x)\right]\frac{\sqrt{\eta}}{2x^{3/2}\sqrt{1-\eta x}}\ ,
\end{equation}
where $\varphi_\pm(x)=p_A\left(\lambda\pm \sqrt{\eta(1/x-\eta)}\right)$.
The distribution displays some interesting features that do not depend on the specific form of $p_A$. Its typical form
is plotted in Fig.~\ref{fig.1}.
It has a peak
of height $\mathcal{O}(1/\eta)$ at $\im G_{ii}\approx \eta$;
for large $\im G_{ii}\ll \eta^{-1}$ it has a power law decay
\begin{equation}
 p_{\im G}(x)\propto p_A(\lambda) \eta^{1/2} x^{-3/2}
\end{equation}
with a cutoff at $x=\im G_{ii}=\eta^{-1}$, where it diverges as $(\eta^{-1}-x)^{-1/2}$.
Note that the $\eta\to 0^+$ limit of the distribution is singular and must be taken after the integration
when computing expectation values. These features are typical of localised phases, see, \emph{e.g.},~\cite{Mirlin1994,Tarquini2016}. 
In order to unveil that for $\gamma \in (1,2)$ the system is instead delocalised but non-ergodic, one 
has to study the statistics of $\im G_{ii}(\lambda-\imi\eta)$ with the scaling $\eta\approx N^{-\delta}$.

Looking at eq.~\eqref{eq:img1}, we foresee three possible behaviours, with a critical
value $\eta_c=N^{1-\gamma}$ discriminating between them.
\begin{itemize}
 \item If  $\eta\gg \eta_c$ ($\delta>\gamma-1$) then $\eta$ dominates the numerator of~\eqref{eq:img1}
 and the previous discussion still holds.
 \item If 
 $\eta\ll\eta_c$ ($\delta<\gamma-1$) then the $\im\overline{G}$ term dominates.
 For finite large $N$ the previous arguments still work, but with $\mu^2 N^{1-\gamma}\im \overline{G}$
 replacing $\eta$.
 \item In the critical case $\eta=\eta_c$, the two terms are of the same order and they both contribute to
 the quantity setting the scales for the probability distribution.
\end{itemize}
\noindent
A similar treatment yields the statistics of $\re G_{ii}(z)$. The result is qualitatively
similar to the imaginary part, but more involved and perhaps less instructive.
For the real part, the limit $\eta\to 0^+$ is not singular and its typical value is of order one.
The typical plot
for $p_{\mathrm{Re} G}(x)$ is shown in Fig.~\ref{fig.1} (right).

In summary, we find that the usual scaling ($N\rightarrow \infty$ first and $\eta\rightarrow 0$ later) is blind to the
existence of the non-ergodic delocalised phase, whose existence can be instead revealed focusing on $\eta=1/N^\delta$. 
For any $\delta<1$, in a standard localised phase the typical value of $\im G_{ii}(x)$ is always of the order of $\eta$, whereas in a standard delocalised phase the typical value of $\im G_{ii}(x)$ tends to a finite value. The behaviour in the non-ergodic delocalised phase is intermediate between these two cases:  the typical value of $\im G_{ii}(x)$ decreases with $\eta$, as it would happen in a localised phase, but only until the value $\eta_c$ is reached. For $\eta\ll \eta_c$ it remains of the order of $\eta_c$, as it would happen in a delocalised phase (with the important difference that $\eta_c$ is not of order one but vanishes as $N^{1-\gamma}$). 
In the next section we relate this result to the scaling of the eigenstate components
using the Dyson Brownian motion technique.

\section{Dyson Brownian motion}
The Dyson Brownian motion (DBM) is a matrix-valued stochastic process in which each element of
the matrix undergoes an independent Brownian motion,
\begin{equation}\label{eq:DBM}
  \di M_{ij}(t)=\sqrt{\frac{(1+\delta_{ij})\sigma^2}{2}} \di W_{ij},
\end{equation}
where $W_{ij}$ are independent standard complex Wiener processes, with $W_{ij}=W^*_{ji}$.
With initial conditions $M(0)=0$, $M(t)$ is a random GUE matrix with variance $\sigma^2t$.
This technique was in fact introduced by Dyson to study spectral properties of
the Gaussian invariant ensembles~\cite{Dyson1962}.

If we consider instead initial conditions $M(0)=A$ and set $\sigma^2=N^{-\gamma}$,
at $t=\mu^2$ we obtain $M(\mu^2)=H$, the Hamiltonian~\eqref{eq:GRP} of the generalised
Rosenzweig-Porter model.
Using perturbation theory on a discretised version of~\eqref{eq:DBM}
(or equivalently It\=o calculus), stochastic differential equations
can be derived,
describing the evolution of the eigenvalues and eigenvectors under the Brownian evolution.
Denoting the $i$-th eigenvalue (sorted in increasing order) by $\lambda_i$ and the corresponding eigenvector
$\psi^{(i)}=(\psi^{(i)}_1,\dots,\psi^{(i)}_N)$, we obtain the stochastic differential equations \cite{Bourgade2013,Allez2015}

\begin{equation}\label{eq:SDEeval}
\di \lambda_i =
\frac{1}{N^{\gamma}}\sum_{j\neq i} \frac{1}{\lambda_i-\lambda_j}\di t+ \frac{1}{N^{\gamma/2}}\di b_i\ ,
\end{equation}
\begin{equation}
\label{eq:SDEevec}
\di \psi^{(i)} = -\left[\sum_{j\neq i}\frac{1}{(\lambda_i-\lambda_j)^2}\right]
\frac{\psi^{(i)}}{2N^{\gamma}} \di t+ \frac{1}{N^{\gamma/2}}
\sum_{j\neq i}\frac{\psi^{(j)} \di b_{ij}}{\lambda_i-\lambda_j}\ ,
\end{equation} 
with initial conditions $\lambda_i(0)=a_i$, $\psi^{(i)}_j(0)=\delta_{ij}$.
The noise terms are real ($b_i$) and complex ($b_{ij}=b_{ji}^*$)
standard Wiener processes.

It is also useful to establish the equation verified by the following quantities \cite{Bourgade2013,Allez2015}
\begin{equation}
 u_{i|j}=[|\psi_j^{(i)}|^2]\ ,
\end{equation}
where $[\cdot\cdot\cdot]$ indicates the
average over the eigenvector noise $b_{ij}$. Since the evolution of the eigenvalues is decoupled from that of the eigenvectors, the average $[\cdots]$ 
does not affect the eigenvalues.  
Using It\=o's calculus one obtains for a fixed realisation of the eigenvalues the evolution equation \cite{Allez2015}
\begin{equation}\label{eq:uEvolution}
 \partial_t u_{i|j}(t) = N^{-\gamma}\sum_{k\neq i} \frac{u_{k|j}(t)-u_{i|j}(t)}{(\lambda_k-\lambda_i)^2}\ ,
\end{equation}
with initial conditions $u_{i|j}=\delta_{ij}$.

\section{Resolvent and DBM}
We first show that the DBM provides an alternative way to find the previous results on the resolvent for $\gamma \in (1,2)$. 
To do so, we first consider the (non-averaged) eigenvalue density
\begin{equation}\label{eq:rho}
\rho(\lambda,t)=\frac{1}{N}\sum_i\delta(\lambda-\lambda_i(t))\ ,
\end{equation}
and its Stieltjes transform, the resolvent
\begin{equation}\label{eq:G-tr-def}
G(z,t)=\int \frac{\rho(\lambda,t)}{z-\lambda} \di\lambda=\frac{1}{N}\tr\resMat(z,t)\ .
\end{equation}
Dean's equation~\cite{Dean1996} provides a way to derive from the SDE for the eigenvalues~\eqref{eq:SDEeval}
a closed stochastic evolution equation for $\rho(\lambda,t)$. A Stieltjes transformation then
gives a closed equation for $G(z,t)$, which is a stochastic (complex) Burgers' equation
\begin{equation}\label{eq:G-Burgers-Dean}
\partial_t G(z,t) =
- \frac{1}{N^{\gamma-1}} G(z, t) \partial_z G(z, t) + \frac{1}{N^{\gamma/2}}\bar{\eta}(z,t)\ ,
\end{equation}
where $\bar{\eta}$ is an order one, Gaussian noise with
\begin{equation}
\braket{\bar{\eta}(z,t)\bar{\eta}(z',t')}
= - \delta(t-t')\partial_z \partial_{z'} \frac{\braket{G(z,t)-G(z',t')}}{z-z'}\ .
\end{equation}
Note that $G$ can be written explicitly as a function of $\{\lambda_i\}$, so its evolution equation
can be derived directly from eq.~\eqref{eq:SDEeval} using It\=o's lemma~\cite{Allez2015}. The result is again a stochastic
Burgers' equation, however the noise term appears in a less appealing form.
While still quite complicated, the form~\eqref{eq:G-Burgers-Dean} clarifies
what the order in $N$ of each term is, and the transitions at $\gamma=1$ and $\gamma=2$ appear
naturally in it. \\
In the intermediate phase the leading term is 0, \textit{i.e.}
$G(z,t)\approx G(z,0)= G_A(z)$ at all $t$, as found previously. The first correction gives a deterministic
inviscid Burgers' equation, well known in random matrix
theory\footnote{The treatment of the Dyson Brownian motion is normally applied to models corresponding to $\gamma=1$~\cite{Rogers1993,Allez2015,Blaizot2010}.
    The standard Burgers' equation is recovered with the rescaling $t\rightarrow N^{\gamma-1}t$.}~\cite{Rogers1993,Blaizot2010}.\\
We can then focus on the local resolvent, averaged over the off-diagonal noise $W_{ij}$ only \cite{Allez2015}:  
\begin{equation}\label{eq:locres}
 U_j(z,t)=[G_{jj}(z,t)] =\sum_i \frac{u_{i|j}}{z-\lambda_i(t)}\ .
\end{equation}
The evolution equation for $U_j$ is derived from equations~(\ref{eq:SDEeval},\ref{eq:uEvolution})
again using It\=o's lemma. The resulting stochastic equation has a structure similar to~\eqref{eq:G-Burgers-Dean}.
Keeping only the leading term in $N$ for $\gamma \in (1,2)$, we obtain the evolution equation
\begin{equation}\label{eq:U-deterministic}
\partial_t U_j(z, t) = - \frac{1}{N^{\gamma-1}} G_A(z) \partial_z U_j(z, t)\ .
\end{equation}
The evolution is deterministic, and the randomness in $U_j$ comes only from the
initial condition $U_j(z,0)=1/(z-a_j)$. The solution to~\eqref{eq:U-deterministic} is
\begin{equation}
  U_j(z,t)^{-1} = z - a_j - \frac{t}{N^{\gamma-1}}G_A(z)\ .
\end{equation}
Comparing with eqs.~(\ref{eq:reg1},\ref{eq:img1}), the result evaluated at $t=\mu^2$ coincides with what we obtained
for the local resolvent $G_{ii}$ from the recurrence equation~\eqref{eq:Grecurrence}.
With that technique it is not necessary to take the $[\dots]$-expected value, which means that $G_{ii}$
is self-averaging for large $N$ with respect to the $W_{ij}$s.

\section{Eigenvectors delocalisation and statistics of the local resolvent}
In order to understand the amount of delocalisation of the eigenvectors we  focus on the solution of 
~\eqref{eq:uEvolution} at $t=\mu^2$.
To extract information on the region of the Hilbert space over which the eigenstates
are delocalised, we consider the following ansatz,
inspired by the results on the statistics of the local resolvent and of Ref.~\cite{Kravtsov2015}.
Assume that $u_{i|j}$ is of order $N^{-\alpha}$ for $|i-j|\approx N^\alpha$ and much smaller on the remaining
$\approx N-N^{\alpha}$ sites. Then the sum in~\eqref{eq:uEvolution} has $N^\alpha$ contributions,
each of which is of order $N^{-2\alpha+2}$
because $\lambda_i-\lambda_j\approx N^{\alpha-1}$. Hence
\begin{equation}
\partial_t u_{i|j} \approx \frac{1}{N^{\gamma}}
N^{\alpha}\cdot N^{-\alpha} N^{2-2\alpha} \overset{!}{\approx} N^{-\alpha}\ ,
\end{equation}
showing that the ansatz is consistent only if $\alpha = 2-\gamma$ and thus establishing that 
the eigenvectors are delocalised but only on $N^{2-\gamma}$ sites.
This result supports the picture that going from the localised to the intermediate phase
the states spread from a single site to $\approx N^{2-\gamma}$ states closest in energy,
and is compatible with the result of Ref.~\cite{Kravtsov2015} for the fractal
dimension of the eigenstates.
Correspondingly, the eigenvalues are correlated if their distance is of order $N^{1-\gamma}$ or less
but become uncorrelated on larger scales. Hence one expects Poisson-like statistics on larger scales 
in agreement with the results found in Ref.~\cite{Kravtsov2015}.

\begin{figure}
    \begin{center}
        \includegraphics[width=\linewidth]{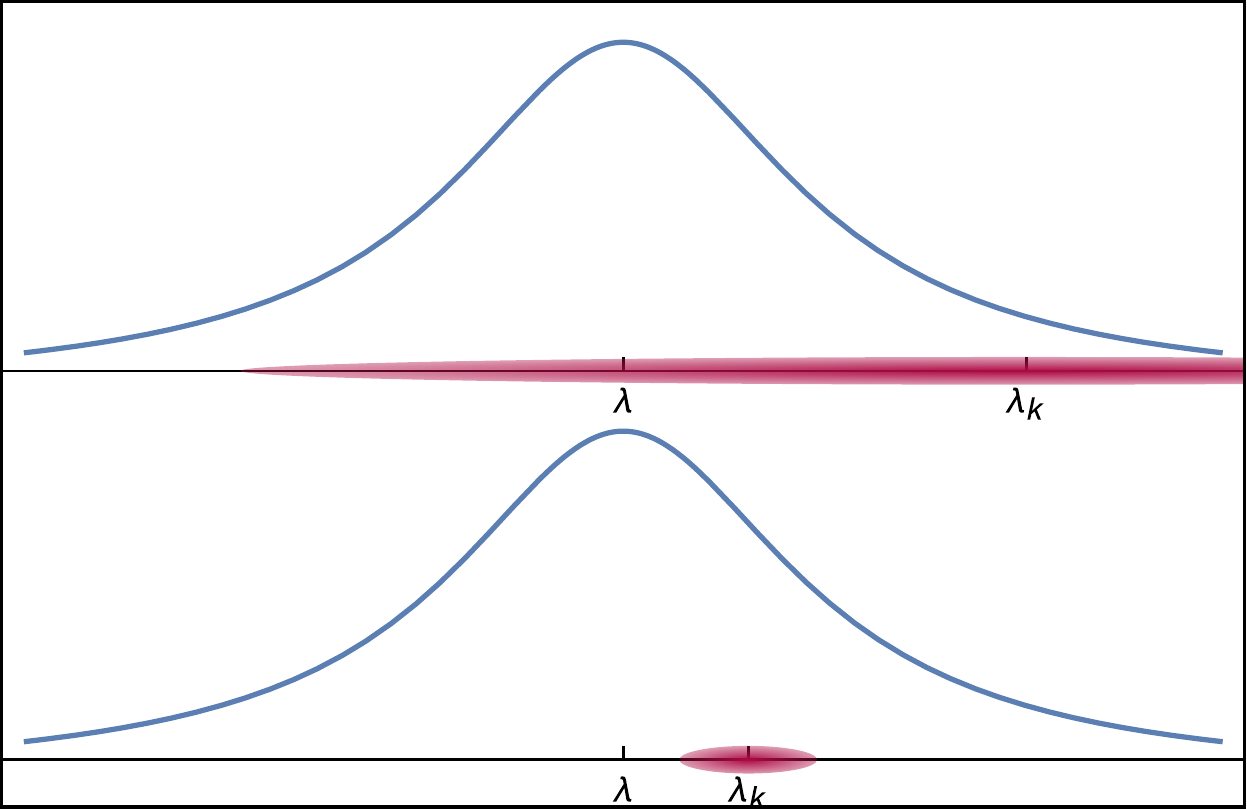}
        \caption{Representation of the packet picture for the contributions to $\im G_{ii}$, see eq.~\eqref{eq:G-packet}.
            The shaded area represents the packet associated to the level $\lambda_k$, of width $\approx\eta_c=N^{1-\gamma}$.
            The width of the Lorentzian is $\eta$. If $\eta>\eta_c$ (lower panel) all the levels in the packet contribute
            approximatively the same. If $\eta<\eta_c$ (upper panel) this is no longer the case.}
        \label{fig.2}
    \end{center}
\end{figure}

We now show that these insights provide a natural explanation of the statistics of the local resolvent found previously. 
The imaginary part of the local resolvent is given by the sum 
\begin{equation}\label{eq:G-packet}
  \im G_{ii}(\lambda-\mathrm{i} \eta)=\sum_k\frac{|\psi_k^{(i)}|^2 \eta}{(\lambda-\lambda_k)^2+\eta^2}\ .
\end{equation}
The $k$-th level's contribution to the sum is given by two factors: the eigenvector weight
$|\psi_k^{(i)}|^2$,
    and a weight depending on $\lambda_k$ as a Lorentzian centred in $\lambda$.
    In the completely localised phase ($\gamma>2$) $|\psi_k^{(i)}|^2 =\delta_{ki}$,
    so only $\lambda_i$ contributes.
    For large $N$, $\lambda_i\approx a_i$ is drawn randomly from $p_A$. If $\lambda$ is in the bulk of the spectrum,
    typically $|\lambda-\lambda_i|\approx 1$, giving a peak
    of the distribution at $\im G_{ii} \approx \eta$.
    If instead $|\lambda-\lambda_i|\lesssim \eta$, which happens with probability $\propto p_A(\lambda)\eta$,
    then
    $\im G_{ii}\lesssim 1/\eta$. These are the rare fluctuations that populate the tail of
    the distribution close to the cutoff at $1/\eta$.
    In between there is the regime $1\ll |\lambda-\lambda_i| \ll \sqrt{\eta}$, where the density of states
    is approximately constant, corresponding to the power law $\propto p_A(\lambda) \im G^{-3/2}$. \\
    We now turn to our region of interest: the delocalised non-ergodic phase $1<\gamma<2$.
    From the the previous discussion on eigenvector delocalisation, we know that the leading contribution comes from
    $\approx N^{2-\gamma}$ levels, which come in ``packets'' (or mini-bands)
    of eigenvalues of width $\eta_c= N^{1-\gamma}$, centred around $\lambda_i$.
    For large enough $N$ and $\eta_c\ll \eta \ll 1$, the width of the packet is much smaller than that of the
    Lorentzian, so the packet behaves coherently,
    with all the eigenvalues in the packet being close to $\lambda$ if and only if $\lambda_k$ is, and so on (see Fig.~\ref{fig.2} - lower panel).
    The total contribution from the packet is identical to the single-level contribution
    in the localised phase. In particular, the typical value of the local resolvent is of the order of $\eta$. 
    This explains the result we found for $\eta_c\ll \eta \ll 1$, see eq.~\eqref{eq:img-pdf}.\\
    The picture changes when the width of the Lorentzian becomes smaller than that of the
    eigenvalue packet. In this case, the probability that the Lorentzian overlaps with the packet is of order $\eta_c$. 
    When this happens, the sum over $k$ in \eqref{eq:G-packet} is of the order of $1/\eta_c$ independently of the 
    value of $\eta$ as long as $\eta>1/N$.
    States which are $O(1)$ away 
     from the centre of the Lorentzian are only important in determining
     the typical value of $\im G_{ii}$: their weights 
     are of order $N^{-\gamma}$ \cite{Kravtsov2015,Bourgade2013} and hence 
     their overall contribution leads to the result $\im G_{typ}\sim N^{1-\gamma}=\eta_c$
     found previously.
     
 \section{Conclusion}
We investigated the localisation properties of the generalised Rosenzweig-Porter model,
using a recurrence relation for the local resolvent and the Dyson Brownian motion. Our main focus was 
the non-ergodic delocalised phase unveiled in Ref.~\cite{Kravtsov2015}, and of which we confirmed 
the existence using complementary techniques. 
Interpreting the model as the combination of on-site random energies $a_i$ and a structurally disordered
hopping, we found that each eigenstate is delocalised over $N^{2-\gamma}$ sites close in energy
$|a_j-a_i|\leq N^{1-\gamma}$, in agreement with the fractal properties found
in Ref.~\cite{Kravtsov2015}.\\
The other main result of our work is the characterisation of the statistics of the local resolvent in the 
non-ergodic delocalised phase. In particular, we showed that its existence can be revealed studying a
non-standard scaling limit in which the small additional imaginary part $\eta$ vanishes as $1/N^\delta$. 
The value $\eta_c$ at which the statistics displays a cross-over from a behaviour characteristic of standard localised phases to a behaviour similar to the one of standard delocalised phases is equal to the typical level spacing, $1/N$, times the number of sites, $N^{2-\gamma}$, over which the eigenvectors are delocalised. Thus, from the local resolvent 
statistics one has a direct access to the non-ergodic properties of the delocalised phase. 
After the completion of this work, we became aware of \cite{Altshuler2016} in which the statistics of the local resolvent in a non-standard scaling limit 
is also proposed and used to probe the existence of a delocalised non-ergodic phase. However, 
the type of cross-over and of non-ergodic delocalised phase are different from the ones studied in this work. In the case studied in  \cite{Altshuler2016}, the delocalised non-ergodic phase should have a typical imaginary part of the local resolvent that does not vanish in the large $N$ limit, moreover below a cross-over scale $\eta_c$ the local resolvent, i.e. the local density of states, should cease to be a smooth function. This is a distinct cross-over from the one found in our work, signalling that the two non-ergodic delocalised phases are different.
\acknowledgements
We gratefully acknowledge insightful discussions with Joe Bhaseen, Jean-Philippe Bouchaud, Paul Bourgade,
Yan Fyodorov, Alessandro Silva and
Marco Tarzia.
DF is supported by the EPSRC Centre for Doctoral
Training in Cross-Disciplinary Approaches to Non-Equilibrium Systems
(CANES, EP/L015854/1). GB is partially supported by the NPRGGLASS ERC grant and by a grant from the Simons Foundation (\#454935, Giulio Biroli). 

\bibliography{nonergRM}

\begin{thebibliography}{10}

\bibitem{Eisert2015}
J.~Eisert, M.~Friesdorf, and C.~Gogolin,
\newblock Nature Physics {\bf 11}, 124 (2015).

\bibitem{BaskoAleinerAltshuler}
D.~Basko, I.~Aleiner, and B.~Altshuler,
\newblock Annals of Physics {\bf 321}, 1126  (2006).

\bibitem{Altshuler1997}
B.~L. Altshuler, Y.~Gefen, A.~Kamenev, and L.~S. Levitov,
\newblock Phys. Rev. Lett. {\bf 78}, 2803 (1997).

\bibitem{Biroli2012}
G.~Biroli, A.~C. Ribeiro-Teixeira, and M.~Tarzia,
\newblock arXiv:1211.7334.

\bibitem{DeLuca2014a}
A.~{De Luca}, A.~Scardicchio, V.~E. Kravtsov, and B.~L. Altshuler,
\newblock arXiv:1401.0019.

\bibitem{DeLuca2014b}
A.~De~Luca, B.~L. Altshuler, V.~E. Kravtsov, and A.~Scardicchio,
\newblock Phys. Rev. Lett. {\bf 113}, 046806 (2014).

\bibitem{Nandkishore15}
R.~Nandkishore and D.~A. Huse,
\newblock Annual Review of Condensed Matter Physics {\bf 6}, 15 (2015).

\bibitem{Pino2015}
M.~Pino, L.~B. Ioffe, and B.~L. Altshuler,
\newblock Proc. Natl. Acad. Sci. U.S.A. {\bf 113}, 536 (2016).

\bibitem{Goold2015}
J.~Goold {\em et~al.},
\newblock Phys. Rev. B {\bf 92}, 180202 (2015).

\bibitem{Altshuler2016}
B.~Altshuler, E.~Cuevas, L.B.Ioffe, and V.E.Kravtsov,
\newblock arXiv:1605.02295.

\bibitem{Tikhonov2016a}
K.~Tikhonov, A.~Mirlin, and M.~Skvortsov,
\newblock arXiv:1604.05353.

\bibitem{Tikhonov2016b}
K.~Tikhonov and A.~Mirlin,
\newblock arXiv:1608.00331.

\bibitem{BarLev2015}
Y.~Bar~Lev, G.~Cohen, and D.~R. Reichman,
\newblock Phys. Rev. Lett. {\bf 114}, 100601 (2015).

\bibitem{BarLev2016}
Y.~{Bar Lev} and D.~R. Reichman,
\newblock EPL {\bf 113}, 46001 (2016).

\bibitem{Vosk2015}
R.~Vosk, D.~A. Huse, and E.~Altman,
\newblock Phys. Rev. X {\bf 5}, 031032 (2015).

\bibitem{Potter2015}
A.~C. Potter, R.~Vasseur, and S.~A. Parameswaran,
\newblock Phys. Rev. X {\bf 5}, 031033 (2015).

\bibitem{Bloch}
J.-y. Choi {\em et~al.},
\newblock Science {\bf 352}, 1547 (2016).

\bibitem{Kravtsov2015}
V.~E. Kravtsov, I.~M. Khaymovich, E.~Cuevas, and M.~Amini,
\newblock New Journal of Physics {\bf 17}, 122002 (2015).

\bibitem{Oren2010}
I.~Oren and U.~Smilansky,
\newblock Journal of Physics A {\bf 43}, 225205 (2010).

\bibitem{Manning2015}
M.~L. Manning and A.~J. Liu,
\newblock EPL {\bf 109}, 36002 (2015).

\bibitem{Cao2016}
X.~Cao, A.~Rosso, J.-P. Bouchaud, and P.~L. Doussal,
\newblock arXiv:1607.04173.

\bibitem{bourgadeprivatecomm}
P.~Bourgade,
\newblock private communication.

\bibitem{RosenzweigPorter}
N.~Rosenzweig and C.~E. Porter,
\newblock Phys. Rev. {\bf 120}, 1698 (1960).

\bibitem{Brezin1996}
E.~Br\'{e}zin and S.~Hikami,
\newblock Nucl. Phys. B {\bf 479}, 697  (1996).

\bibitem{Kunz1998}
H.~Kunz and B.~Shapiro,
\newblock Phys. Rev. E {\bf 58}, 400 (1998).

\bibitem{Pandey1995}
A.~Pandey,
\newblock Chaos, Solitons \& Fractals {\bf 5}, 1275  (1995).

\bibitem{Altland1997}
A.~Altland, M.~Janssen, and B.~Shapiro,
\newblock Phys. Rev. E {\bf 56}, 1471 (1997).

\bibitem{Guhr1996}
T.~Guhr,
\newblock Annals of Physics {\bf 250}, 145  (1996).

\bibitem{Shukla2016}
P.~Shukla,
\newblock New Journal of Physics {\bf 18}, 021004 (2016).

\bibitem{Tarquini2016}
E.~Tarquini, G.~Biroli, and M.~Tarzia,
\newblock Phys. Rev. Lett. {\bf 116}, 010601 (2016).

\bibitem{Mirlin1994}
A.~D. Mirlin and Y.~V. Fyodorov,
\newblock Phys. Rev. Lett. {\bf 72}, 526 (1994).

\bibitem{Dyson1962}
F.~J. Dyson,
\newblock Journal of Mathematical Physics {\bf 3} (1962).

\bibitem{Bourgade2013}
P.~Bourgade and H.-T. Yau,
\newblock arXiv:1312.1301.

\bibitem{Allez2015}
R.~Allez, J.~Bun, and J.-P. Bouchaud,
\newblock arXiv:1412.7108.

\bibitem{Dean1996}
D.~S. Dean,
\newblock Journal of Physics A {\bf 29}, L613 (1996).

\bibitem{Rogers1993}
L.~C.~G. Rogers and Z.~Shi,
\newblock Probability Theory and Related Fields {\bf 95}, 555 (1993).

\bibitem{Blaizot2010}
J.-P. Blaizot and M.~A. Nowak,
\newblock Phys. Rev. E {\bf 82}, 051115 (2010).

\end{thebibliography}
\bibliographystyle{h-physrev}

\end{document}